# Thermal transport of amorphous carbon and boron-nitride monolayers

Yu-Tian Zhang (张雨田)[1], Yun-Peng Wang (王云鹏)[2], Yu-Yang Zhang (张余洋)[1,3†], Shixuan Du (杜世萱)[1,4†], Sokrates T. Pantelides[5,1†]

1 University of Chinese Academy of Sciences and Institute of Physics, Chinese Academy of Sciences, Beijing 100049, China
2 Hunan Key Laboratory for Super Microstructure and Ultrafast Process, School of Physics and Electronics, Central South University, Changsha 410083, China
3 CAS Center for Excellence in Topological Quantum Computation, University of Chinese Academy of Sciences, Beijing 100049, China
4 Songshan Lake Materials Laboratory, Dongguan, Guangdong 523808, China
5 Department of Physics and Astronomy and Department of Electrical and Computer Engineering, Vanderbilt University, Nashville, Tennessee 37235, USA

## Abstract

Two-dimensional (2D) materials like graphene and h-BN usually show high thermal conductivity, which enables rich applications in thermal dissipation and nanodevices. Disorder, on the other hand, is often present in 2D materials. Structural disorder induces localization of electrons and phonons and alters the electronic, mechanical, thermal, and magnetic properties. Here we calculate the in-plane thermal conductivity of both monolayer carbon and monolayer boron nitride in the amorphous form, by reverse nonequilibrium molecular dynamics simulations. We find that the thermal conductivity of both monolayer amorphous carbon (MAC) and monolayer amorphous boron nitride (ma-BN) are about two orders of magnitude smaller than their crystalline counterparts. Moreover, the ultralow thermal conductivity is independent of the temperature due to the extremely short phonon mean free path in these amorphous materials. The relation between the structure disorder and the reduction of the thermal conductivity is analyzed in terms of the vibrational density of states and the participation ratio. ma-BN shows strong vibrational localization across the frequency range, while MAC exhibits a unique extended *G'* mode at high frequency due to its *sp²* hybridization and the broken $E_{2g}$ symmetry. The present results pave the way for potential applications of MAC and ma-BN in thermal management.

## Introduction

Two-dimensional materials show diverse thermal conductivities. For example, graphene (Gr) and monolayer h-BN possess very high thermal conductivity ($\kappa$ = 2000 – 5000 Wm$^{-1}$K$^{-1}$ for graphene [1-4] and ~750 Wm$^{-1}$K$^{-1}$ for h-BN [5,6]). These values make the materials useful in heat-removal applications in nanodevices [7-12]. Meanwhile, some 2D materials, like SnSe, 2D tellurium, selinene, stanene, and $MoO_3$, exhibit a low κ, which is desirable for thermoelectric devices and thermal protective barriers in nano- and micro-scale devices [13-16].

Disorder, which is often present in materials, significantly affects the electronic, mechanical, and thermal properties [17-22]. Amorphousness, which can be viewed as an extreme limit of disorder, modulates the vibrational modes of materials and thus affects the thermal conductivity. There have been several theoretical studies on the thermal conductivity of amorphous graphene (a-Gr) and amorphous 2D boron nitride (a-BN) by using classical molecular dynamics simulations [19,23-25]. However, in these papers, the atomic structures of a-Gr and a-BN were constructed by intentionally introducing excessive point defects into pristine graphene and h-BN lattice. Such an amorphization process can, in principle, produce either a totally pure Zachariasen continuous random network (Z-CRN) [26], while recent experiments have found that monolayer amorphous carbon (MAC) indeed contains nanocrystallites [27]; Similarly, the process may lead to highly defective h-BN rather than truly a-BN. The thermal properties of amorphous counterparts of Gr and h-BN with realistic atomic structures are of both fundamental and practical value but are still largely unknown.

In this work, motivated by the recent synthesis of MAC and our prediction of monolayer amorphous BN (ma-BN) [27,28], we performed a systematic study of the thermal transport of these two amorphous materials. By using reverse nonequilibrium molecular dynamics (RNEMD) simulations, we calculated the in-plane thermal conductivity of crystallite form of MAC [27] and ma-BN with a Z-CRN configuration [28]. Compared with their crystalline counterparts, which possess high thermal conductivities and prominent temperature dependence, we found that the thermal conductivities of the amorphous phases of these two materials are significantly reduced and are insensitive to temperature. To better understand the origin of the reduction in thermal conductivity, we further analyzed the phonon properties including the vibrational density of states and the participation ratio. It is found that the characteristic peaks in the vibrational density of states of Gr and h-BN are weakened and broadened in the amorphous structures, which reflects the reduction of phonon lifetimes. Moreover, the participation ratio of both the MAC and the ma-BN are overall much lower than that of crystalline Gr and h-BN, indicating significant localization of vibrational modes. Specifically, ma-BN shows strong localization of vibrational modes over the whole frequency range, because of the pure CRN atomic structure. By contrast, due to the existence of substantial crystallites in MAC, vibrational modes are fairly extended at several frequency ranges. Due to the $sp^2$-hybridized carbon network and the absence of $E_{2g}$ symmetry caused by amorphousness, MAC exhibits a unique high-frequency extended *G'* mode that resembles the *G* mode of Gr.

# Methods

We took the atomic structures of MAC and ma-BN from our previous work [27,28], which were generated by kinetic Monte Carlo (kMC) simulations. The initial configurations comprise atoms at random positions and the kMC simulates the annealing process at low temperature. We chose representative snapshots of MAC (CRN with nanocrystallites, as shown in Fig. 1a) and ma-BN (CRN with *pseudocrystallites*, as shown in Fig. 1b). The in-plane sizes of the periodic cells are both 45 Å × 45 Å, and the monolayers are separated in the z-direction by a 20-Å vacuum layer. To describe the interatomic interactions, we employed the optimized Tersoff potential developed by Lindsay and Broido [29] for Gr and MAC, and the extended Tersoff potential for BN (BN-ExTeP [30]) for h-BN and ma-BN. Thermal conductivity was calculated by RNEMD, which is based on the Muller-Plath algorithm [31]. All the calculations were conducted using the molecular dynamics code LAMMPS [32]. Periodic boundary conditions are employed for MD simulations. A representative supercell of MAC is shown in Fig. 1c, where the heat-flow direction is along the *x* coordinate. For MAC and ma-BN, the size effects are tested by comparing the radial distribution functions with varying sample sizes, as shown in Extended Fig. 1. We also tested the length and width effects in Extended Fig. 2, the results show that 30 nm of supercell length and 4.5 nm of width are sufficient to converge the thermal conductivity. We thereby chose the supercells with size of 30 nm × 4.5 nm. For Gr and h-BN, the sample lengths ($L$) range from 5 nm to 200 nm for extrapolations, while the widths are kept at 4.3 nm. The width convergence is also tested in Extended Fig. 3.

In all simulations, the equations of motion are integrated with a time step of 0.2 fs. The initial structures are first relaxed for 500 ps in an isothermal-isobaric (NPT) ensemble using a Nose-Hover thermostat at the target temperature and zero pressure. Then the simulation is switched to a canonical (NVT) ensemble for another 500 ps, and finally a microcanonical (NVE) ensemble for 500 ps for equilibration. After reaching the steady state, we perform another long-time NVE evolution (> 9 ns) for the thermal conductivity calculations. To calculate the temperature profile, the simulation boxes are equally divided into many bins in the x-direction (each bin is ~1 nm long [31]). Heat is then imposed from the middle bin by exchanging the kinetic energies of "hot"-region atoms with "cold"-region atoms. The frequency of kinetic-energy swaps and the sampling time for averaging are deliberately controlled to ensure: (1) the temperature differences of heat source and heat sink are kept in a reasonable range; (2) the temperature gradient is linear in the middle bins. Typical temperature profiles are shown in Extended Fig. 4.

Based on Fourier's law of heat conduction, the thermal conductivity is calculated as $\kappa = J/(2A \cdot \partial T/\partial x)$, where $A$ is the cross-section area perpendicular to the heat flow direction, and the effective thickness is set to 3.3 Å for both carbon and BN structures, which is the interlayer distance in their crystalline forms. For Gr and h-BN, due to their phonon mean free path (MFP) are longer than our simulation supercells, we further perform extrapolation of the thermal conductivity to infinite length by the empirical equation of Schelling et al. [33]:

$$\frac{1}{\kappa_L} = \frac{1}{\kappa_\infty}\left(\frac{\lambda}{L} + 1\right),$$

where $\kappa_\infty$ is the intrinsic thermal conductivity (length-independent) and $\kappa_L$ is the thermal conductivity at a certain length $L$. The effective phonon MFP, $\lambda$, can also be retrieved from the above formula. For MAC and ma-BN, the thermal conductivities converge quickly with increasing supercell length, whereby an extrapolation is unnecessary. For statistical convergence, we perform five independent calculations of MAC and ma-BN from different initial velocity distributions and average the thermal conductivities; for Gr and h-BN, due to the tedious and expensive extrapolations, only the thermal conductivities at 300 K and 600 K are averaged.

The vibrational density of states (VDOS) was calculated using the Fourier transform of the velocity autocorrelation function (VACF) during a 100 ps NVE run [34]:

$$VDOS = \int_0^\infty \frac{\langle v(t) \cdot v(0) \rangle}{v(0) \cdot v(0)} e^{-i\omega t} dt\,,$$

where ω is the phonon frequency, $v$ is the velocity vector, and $t$ is the correlation time. The participation ratio (PR) can be calculated from the local VDOS [35,36]:

$$PR(\omega) = \frac{1}{N} \frac{(\sum_i VDOS_i(\omega)^2)^2}{\sum_i VDOS_i(\omega)^4}\,,$$

where $N$ is the number of atoms, and $VDOS_i$ is the *VDOS* of the *i*th atom. The local VDOS are calculated by the real-space Green's function method [37,38].

**Results and discussion**

The calculated in-plane thermal conductivities are shown in Fig. 2. We begin by considering the thermal conductivities of crystalline materials as a benchmark. As summarized in Table 1, the mean thermal conductivity of crystalline graphene is 2864 Wm$^{-1}$K$^{-1}$ at 300 K, which falls inside the range of previous reports [1-4,39-44]. For monolayer h-BN, the calculated thermal conductivity is 1985 Wm$^{-1}$K$^{-1}$ at 300 K, which is larger than the both experimental [5,6] and equilibrium MD (EMD) results [45]. We attribute this discrepancy to the different methods (EMD *vs.* RNEMD) and potentials (optimized Tersoff *vs.* BN-ExTeP). The reason we are not using the optimized Tersoff potential for h-BN is that the ma-BN will be unstable in the MD. For both Gr and h-BN, the thermal conductivities decrease with increasing temperature, as shown in Fig.2(a). The nearly linear reduction of $\kappa$ indicates shorter phonon MFP, because of the much more intensive phonon scattering at higher temperatures. This trend is qualitatively consistent with experimental data [2,3,5,6]. We must emphasize that the value of crystalline $\kappa$ can vary significantly by different calculation/measurement methods, therefore we focus on the relative comparison between the crystalline and amorphous materials under the same set of calculation parameters.

The calculated thermal conductivities of MAC and ma-BN are shown in Fig. 2b. The averaged κ is only 13.3 $Wm^{-1}K^{-1}$ for MAC and 7.2 $Wm^{-1}K^{-1}$ for ma-BN at 300 K. Both of these values are two orders of magnitude lower than their crystalline counterparts. The dramatic reduction of κ is attributed to the amorphous atomic arrangements, which encumber the formation of delocalized phonons and lead to intrinsically short phonon MFPs. Note that in Extended Fig. 2(a), the length convergence of MAC is much slower than ma-BN, indicating the longer phonon MFP of MAC. It also explains the higher κ of MAC when compared to ma-BN. Different from the linear κ reduction of Gr and h-BN, the κ of MAC and ma-BN are basically temperature independent (only slightly reduction at T > 600 K for ma-BN). The aforementioned intrinsically short phonon MFPs are hardly affected by the variation of temperature, because they are probably shorter than the spatial periodicity of the temperature-induced ripples, whereby the consequent low κ are also immune to temperature changes. The T-independent κ for MAC is consistent with Ref. [24], where the κ of a-Gr is less sensitive to temperature with increasing defect concentration. However, the value and the temperature dependency of κ of MAC are different from Ref. [19], because the atomic structure of a-Gr in their work is Z-CRN generated by melt-from-the-quench method and is thus qualitatively different from the crystallite MAC in this work. The small simulation cell of 1.25 × 2.16 $nm^2$ and the EMD simulations they used may also account for the discrepancy.

The mechanism of heat transport in amorphous materials is fundamentally different from crystalline materials. Because the lack of atomistic periodicity and thus translational symmetry, the q-vector of phonons in crystals is no longer a good quantum number in amorphous materials, thus the phonon quasiparticles are not well-defined in amorphous structures [20]. The widely-adopted theory of heat transport in disordered solids proposed by Allen and Feldman in 1993 [46] categorizes the heat carriers into three types: propagons, diffusons, and locons. The propagons are low-frequency vibrational modes in analog to the wave-like phonons, while diffusons are of mid-frequency and carry the heat diffusively. Locons are high-frequency modes and spatially localized, they contribute negligibly to thermal conductivity. The low κ of MAC and ma-BN indicates that the low-frequency vibrational modes contribute little to the heat transport, when compare to their crystalline analogs, in which the low-frequency ZA modes are the major heat carriers [5,42-44].

To investigate the significant low and temperature-insensitive thermal conductivities of MAC and ma-BN, we studied the vibrational properties. Since the concept of vibrational density of states (VDOS) is applicable in amorphous solids, we calculate VDOS of both crystalline and amorphous phases. As shown in Fig. 3a and 3b, the VDOS of both Gr and h-BN show characteristic peaks, including the high frequency optical modes near 50 THz and ZA/ZO peaks at ~8 and 16 THz, respectively. The lifetime of carriers within these peaks is inversely proportional to the width of the VDOS peaks [7,23,47,48]. Consequently, thermal conductivity is also inversely proportional to the width of the VDOS peak, as the former is proportional to the carrier lifetimes. We observed significant broadening or even disappearance of such characteristic peaks in both MAC and ma-BN, as shown in Fig. 3a and 3b. The peak broadening indicates a decrease of carrier lifetimes as well as the MFP [49], resulting in low thermal conductivities.

To obtain a deeper understanding of the mechanism of the decreased MFP, we investigate the participation ratio (PR), which is defined as the fraction of atoms that participate in a vibrational mode. The PR measures the spatial localization of vibrational modes. A vibrational mode is localized if the PR equals $O(1/N)$, while it is extended if PR is $O(1)$ [20]. As shown in Fig. 4a, the PR of Gr is a constant across the whole frequency range and is equal to 1, reflecting the perfect extension of all the vibrational modes, which is consistent with the highest thermal conductivity of Gr in 2D materials [19]. Similar to the feature in Gr, as shown in Fig. 4b, the PR of h-BN is larger than 0.5 at most frequencies, reflecting the extended vibrational modes; but it also exhibits a certain degree of localization at some frequencies. The lower PR for h-BN contributes to its smaller thermal conductivity compared to Gr.

The calculated PR of amorphous materials show prominent reduction compared with their crystalline analogs, revealing significant localization of the vibrational modes. As shown in Fig.4(b), the PR of ma-BN remains smaller than 0.2 across the whole frequency range, which reflects strong localization. It is not surprising to observe localization in amorphous materials, where the disorder-induced localization is not only limited to electronic states first proposed by Anderson [17], but also applicable to many other waves-like spin and vibrational waves in disordered media [19]. However, at low frequencies of 0 – 10 THz, the low PR reflects highly localized modes, which contradicts the conventional propagon picture of Allen-Feldman formalism. At high frequency range > 40 THz, the PR is close to zero and matches the typical locon behavior [46].

For MAC, however, the PR shows richer behaviors. The low-frequency (0 – 8 THz) PR is close to zero, as shown in Fig.4(a). For better understanding the vibrational behaviors in real space, we visualize the eigenvectors by solving the secular equation of dynamical matrices. As shown in Fig.5 (a), only a fraction of the atoms participates in the collective motions of the low-frequency modes, while in each region the atoms vibrate along the same direction, i.e., the movements are *in-phase*. These collective vibrational modes are probably related to the presence of nanocrystallites in MAC. These highly localized vibrational modes are different from the propagon picture of Allen-Feldman theory and contribute little to heat transport, thus explains the two-orders-of-magnitude reduction of thermal conductivity of MAC. We then analyze the three PR plateaus as highlighted by blue in Fig.4(a). These delocalized vibrational modes (PR > 0.4 can be regarded as delocalized [19]) are unique in MAC, as they are not observed in ma-BN. The first plateau occurs at ~15 THz, which we ascribe to diffusons. As shown in Fig.5(b), many atoms participate the vibration, but the displacement directions are random, i.e., they are *out-of-phase* vibrations. The second plateau at 30 – 40 THz corresponds to mid-frequency diffusons, their vibrational phases are also disordered, as shown in Fig.5(c).

The prominent PR plateau at 45 – 55 THz are distinct from typical high-frequency locon pictures. The frequency range is 1501.0 – 1667.8 cm$^{-1}$, which is similar to the *G* mode of Gr (1580 – 1600 cm$^{-1}$), thus we call it *G'* mode. It is known that *sp$^2$* carbon systems can excite Raman active *G* mode [50], therefore the *G* mode is also active in MAC because its *sp$^2$* carbon network nature. However, disorder and impurity can split the *G* mode into *G* (1583 cm$^{-1}$) and *D'* mode (1620 cm$^{-1}$) [51,52], the *G* mode is further broadened by the presence of small amount of non-*sp$^2$* carbon atoms in MAC. The amorphousness and the slightly non-*sp$^2$* hybridization result in a broadened extended mode as seen in the PR, i.e., the *G'* mode.

The eigenvectors of *G'* mode are shown in Fig.5(d). The vibrational mode is extended in real space, while the atoms are moving with random phases. This is because the amorphous lattice of MAC breaks the *E$_{2g}$* symmetry of Gr, as a result, the regular in-plane bond stretching of graphene *G* mode is replaced by the chaotic *out-of-phase* vibration of *G'* mode. The predicted *G'* mode is also confirmed by previous experimental observations of broadened D and G mode of a-Gr [27,53].

The vibrational eigenvectors of ma-BN are shown in Fig.6, which lack *in-phase* collective movements even at low frequencies. The lower PR and the random vibrations of ma-BN throughout the frequency range may originate from its unique structure of *pseudocrystallite* embedded in CRN, which is generally more disordered than the crystallites in MAC. This is a manifestation that the medium range order can influence the vibrational and thermal properties of amorphous materials.

## Conclusions

In summary, we study the vibrational and thermal properties of MAC and ma-BN comprehensively by RNEMD simulations. We find that the thermal conductivities of MAC and ma-BN are temperature independent and are about two orders of magnitude smaller than their crystalline counterparts. Our analysis of the vibrational DOS indicates that the reduction of the thermal conductivity can be explained by the short lifetime of vibrational modes in amorphous structures. Moreover, the amorphous vibrational modes exhibit localization as shown by the reduced PR. Different from the strong localization of ma-BN across the whole frequency range, MAC exhibits three extended vibrational modes. Two of them are diffusons with random vibrational directions, and the unique high-frequency *G'* mode originates from the *sp$^2$* carbon network and the broken *E$_{2g}$* symmetry of MAC.

2D carbon and BN allotropes can exhibit distinct thermal conductivities. For crystalline phases, graphene and h-BN show high thermal conductivities, which is desirable in heat dissipation applications; for the amorphous counterparts, however, we show that MAC and ma-BN possess intrinsic low thermal conductivities that do not vary with temperature, which are desirable in thermal isolations. In addition to the electrical insulating properties and good stability at room temperature [27,28], MAC and ma-BN are promising and robust candidates in thermal management applications. It is also possible to realize different thermal applications by synthesizing different phases (crystalline *vs.* amorphous) of 2D carbon and BN allotropes at controllable growth conditions.

## Acknowledgment

We acknowledge financial support from National Key R&D program of China (Nos. 2019YFA0308500), National Natural Science Foundation of China (61888102 and 12004439), Strategic Priority Research Program of the Chinese Academy of Sciences (Nos. XDB30000000 and XDB28000000), the K. C. Wong Education Foundation, and the Fundamental Research Funds for the Central Universities. Work at Vanderbilt was funded by the McMinn Endowment.

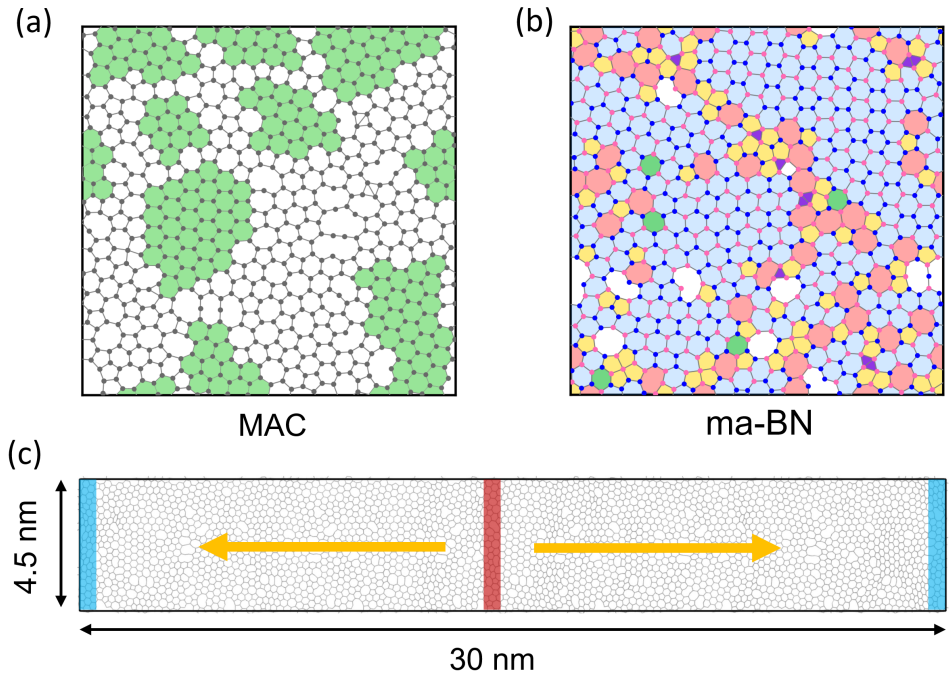

**Fig. 1. Atomic structures of MAC and ma-BN.** (a) and (b) are typical amorphous structures for MAC and ma-BN as reported in Ref. [27, 28]. The green regions are the crystallites of MAC, and the blue regions are the *pseudocrystallites* of ma-BN. Green hexagons are canonical. Note that the *pseudocrystallites* comprise of noncanonical BN hexagons (blue), which maintain the distorted hexagonal lattice, but the on-site elements are random. It is a unique feature of 2D binary amorphous materials. Detailed discussions can be found in Ref. [28]. Pentagon and heptagon are colored in yellow and orange, respectively. (c) A typical supercell of amorphous materials employed in the RNEMD simulation of thermal conductivity. The heat is injected from the middle red bin ("hot" region), the yellow arrows indicate the heat flow directions towards the blue bins, which are the heat sinks ("cold" regions).

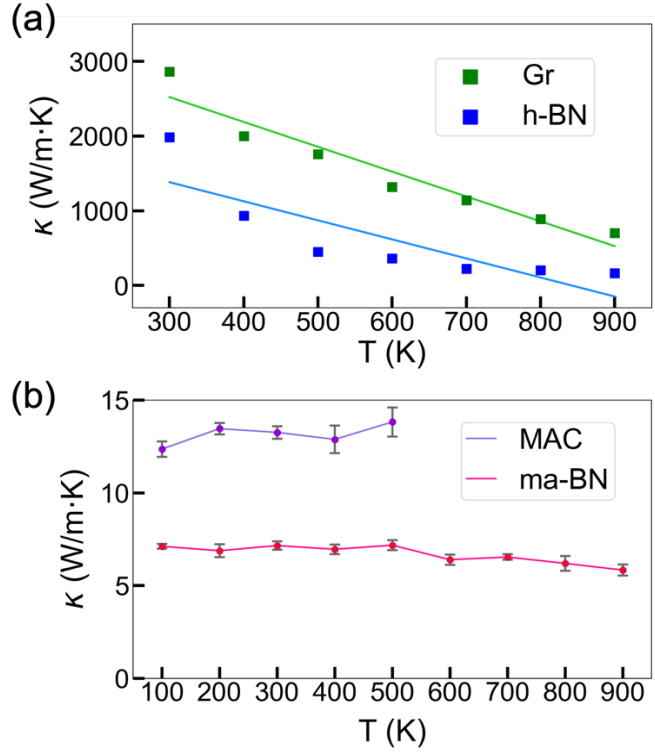

**Fig. 2. Thermal conductivities as a function of temperature.** (a) is for Gr and h-BN. (b) is for MAC and ma-BN. The error bars represent the standard deviations of five independent calculations. Since MAC is unstable at T > 500K, thus the κ is not shown.

**Table 1. Thermal conductivities of Gr and h-BN.** Compared with first-principles calculations, Boltzmann transport equation (BTE), nonequilibrium/equilibrium MD (NEMD/EMD), and experimental measurements.

| κ (300 K) (W/mK) | First-principles | BTE | NEMD/EMD | Exp. | This work |
|---|---|---|---|---|---|
| **Gr** | 2200 [41] | 3500 [42] | 1015 [54] | 3080–5150 [55] | 2864 |
|  | 2897 [43] | 3383 [44] | 3200 [39] | 4840–5300 [1] |  |
| **h-BN** | 800–1600 [56] | 617–1107 [57] | 649 [45] | 751±340 [5,6] | 1985 |

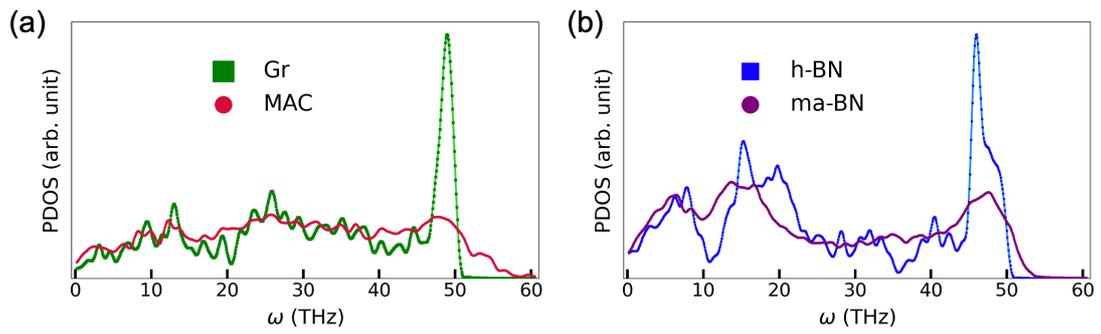

**Fig. 3. The vibrational density of states (VDOS).** (a) is for Gr and MAC. (b) is for h-BN and ma-BN.

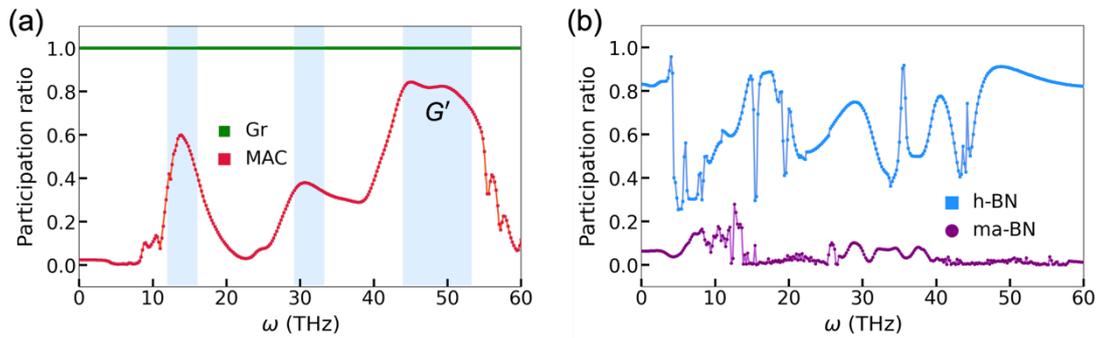

**Fig. 4. The participation ratio (PR).** (a) is for Gr and MAC. (b) is for h-BN and ma-BN. Three extended regions of MAC are highlighted in light blue.

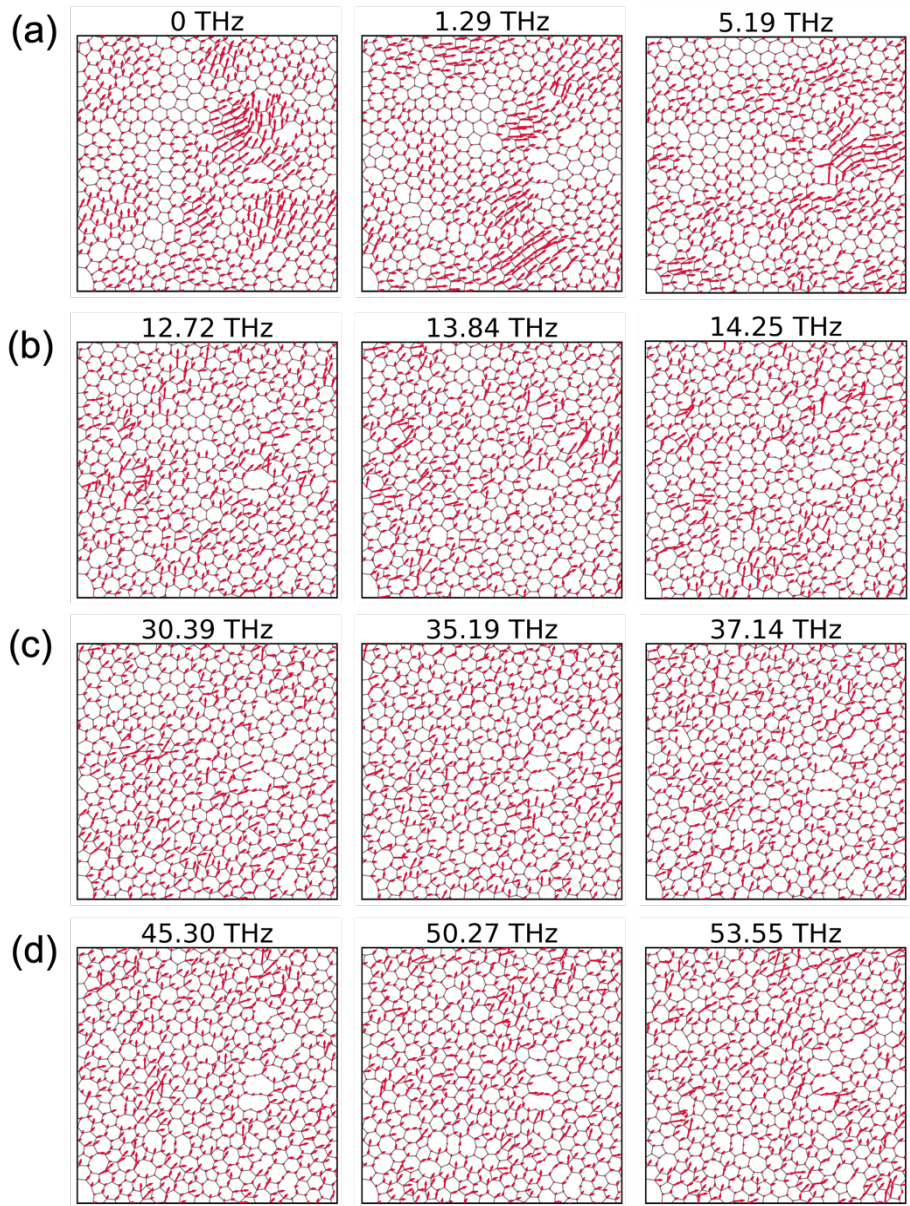

**Fig. 5. Vibrational eigenvectors of MAC.** Representative modes are shown in each frequency region. (a) Three low-frequency localized modes. (b) – (d) correspond to the highlighted regions in Fig. 4(a). (b) Three low-frequency diffusons. (c) Three mid-frequency diffusons. (d) Three high-frequency *G'* modes. Red arrows represent the vibrational directions of each atom.

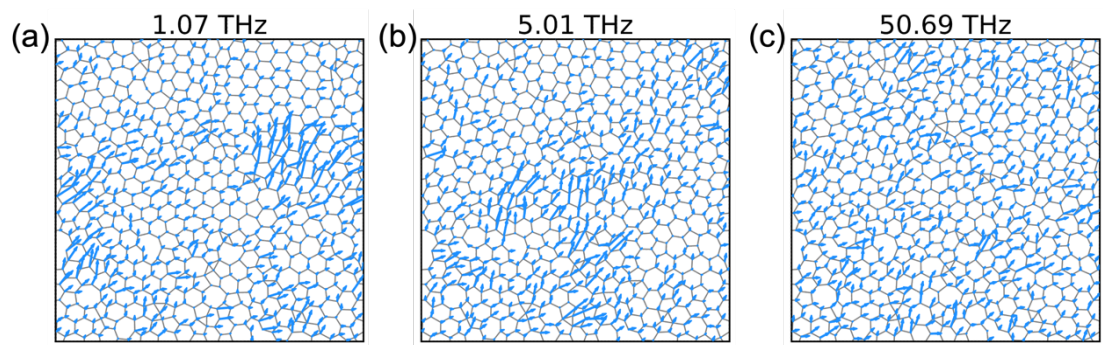

**Fig. 6. Vibrational eigenvectors of ma-BN.** (a) – (c) Vibrational modes at three selected frequencies. Blue arrows represent the vibrational directions of each atom. The atomic movements are *out-of-phase* and are more localized than MAC.

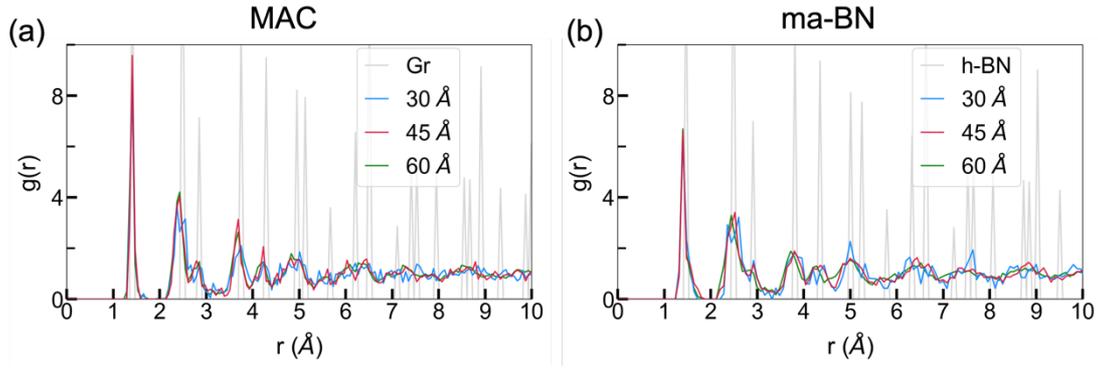

**Extended Fig. 1. Radial distribution functions of (a) MAC and (b) ma-BN with different sizes.** The colored RDF curves correspond to three increasing sizes of unitcell, good convergence are achieved at 45 Å, showing that the structural features are converged and the size effects are negligible.

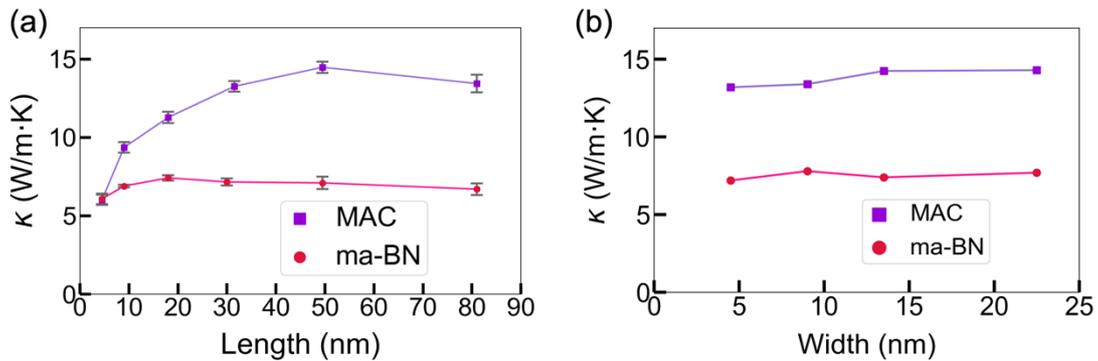

**Extended Fig. 2. Thermal conductivity of MAC and ma-BN with different sizes.** (a) The thermal conductivity as a function of length. The $\kappa$ of ma-BN converge quickly and can be regard as length independent. In contrast, the $\kappa$ of MAC increase with length and peak at ~50 nm. Note that $\kappa$(50 nm) is only 9% larger than $\kappa$(30 nm), thus we take the 30 nm value as the converged $\kappa$ of MAC for all subsequent calculations in this work. (b) The thermal conductivity as a function of width. Both MAC and ma-BN show nearly flat curves of $\kappa$, which means that the 4.5 nm wide unitcells are enough for convergence.

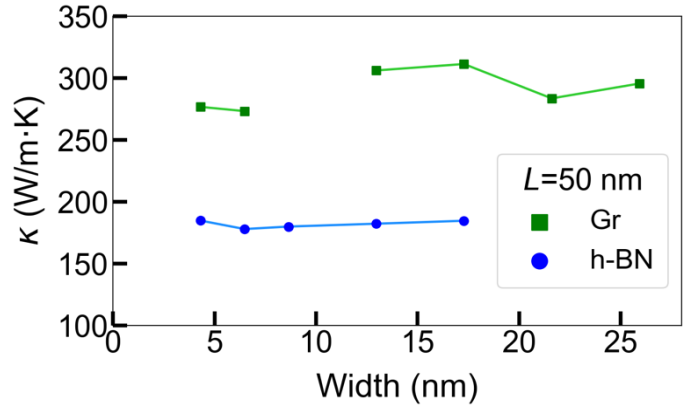

**Extended Fig. 3. Thermal conductivity of Gr and h-BN with different sizes.** The length of the supercells are fixed to 50 nm. The $\kappa$ of h-BN are independent of the width of supercell, while the $\kappa$ of Gr show little fluctuations. For Gr, the $\kappa$ value at 26 nm is only 6.8% larger than that of 4.3 nm, whereby we choose 4.3 nm of supercell width in all subsequent calculations for both Gr and h-BN.

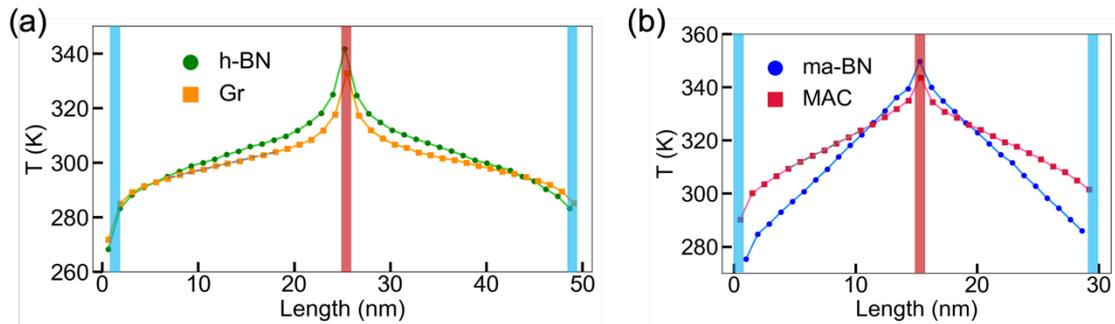

**Extended Fig. 4. Typical temperature profiles in RNEMD simulations.** (a) Temperature gradient of Gr and h-BN at 300 K. (b) Temperature gradient of MAC and ma-BN at 300 K. The middle regions show good linearity, and the temperature differences between hot and cold regions are controlled within reasonable range.